\documentclass[11pt,a4paper]{article}
\usepackage{jcappub}
\usepackage[sort&compress]{natbib}

\begin{document}

\title{Strong gravitational lensing of gravitational waves from double compact binaries --- perspectives for the Einstein Telescope}

\author{Marek Biesiada,$^{1,2}$}
\author{Xuheng Ding,$^1$}
\author{Aleksandra Pi{\'o}rkowska, $^2$}
\author{Zong-Hong Zhu $^1$}

\affiliation{
$^1$ Department of Astronomy, Beijing Normal
University, Beijing 100875, China \\
$^2$ Department of Astrophysics and Cosmology, Institute of Physics,
University of Silesia, Uniwersytecka 4, 40-007 Katowice, Poland

}

\abstract{
Gravitational wave (GW) experiments are entering their advanced stage which should soon
open a new observational window on the Universe. Looking into this future, the Einstein Telescope (ET) was designed to have a fantastic sensitivity
improving significantly over the advanced GW detectors.
One of the most important astrophysical GW sources supposed to be detected by the ET in large numbers are double compact objects (DCO) and some of
such events should be gravitationally lensed by intervening galaxies.

We explore the prospects of observing gravitationally lensed inspiral DCO events in the ET.
This analysis is a significant extension of our previous paper \citet{JCAP_ET}.
We are using the intrinsic merger rates of the whole class of DCO (NS-NS,BH-NS,BH-BH)located at different redshifts as calculated by \citet{BelczynskiII}
by using {\tt StarTrack} population synthesis evolutionary code. We discuss in details predictions from each evolutionary scenario. Our general conclusion is that ET would register about $50-100$ strongly lensed inspiral events per year. Only the scenario in which nascent BHs receive strong kick gives the predictions of a few events per year. Such lensed events would be dominated by the BH-BH merging binary systems. Our results suggest that during a few years of successful operation ET will provide a considerable catalog of strongly lensed events.

 }

\maketitle

\section{Introduction} \label{sec:intro}

Gravitational waves (GW thereafter) are the last of the biggest predictions of the Einstein General Relativity theory which still eludes our direct detection.
There is no doubt that indirect arguments for their existence are extremely strong: from the famous Hulse--Taylor binary pulsar (and other similar systems) to the imprints of primordial gravitational waves on the CMBR recently announced by BICEP2 collaboration \citep{BICEP}. If finally confirmed as a pure uncontaminated primordial signal, B modes in the CMBR would be another indirect proof of the existence of GWs. It is not surprising that already long time ago many teams took the challenge to register GWs from the ground. Technological development of past two decades culminated in the construction of LIGO/VIRGO type interferometric detectors \citep{LIGO,VIRGO} and these experiments are just entering the next, advanced phase. Encouraged by fast qualitative changes which new technologies brought into science there was a commitment of GW scientific community to design and build a new generation detector called the Einstein Telescope (ET thereafter) \citep{Abernathy03}. It will improve an order of magnitude in sensitivity over the advanced laser interferometer detectors LIGO and VIRGO. Moreover its design assumes enlargement of sensitivity range into low (of order of $1Hz$) frequencies. This would have many advantages, but this goal cannot be achieved within one type of the detector. The reason is, that in order to probe low frequencies one needs to face the photon shot-noise so increased power of the laser beam would be beneficial. On the other hand, at high frequencies the main obstacles are radiation pressure and thermal noise, so these two goals contradict each other as far as technology is concerned. Hence the idea is to start with the initial (sensitive to high frequencies) configuration and afterwards complement it with second detectors (one for each of three sets of arms) sensitive to low frequencies. This setting is called ``xylophone'' configuration. We will refer to these two designs further in the paper.

One of the most important astrophysical GW sources supposed to be detected by ET in large numbers are double compact objects (DCO). We are adopting the name used in \citet{BelczynskiII} --- the paper which is central to our study. DCOs comprise of NS-NS, BH-NS and BH-BH binary systems and even though only the first class of DCOs is empirically proven to exist, yet the consistency reasons based on stellar evolution theory lead us to believe in the existence of other classes and moreover encourage us to predict their properties (inspiral rates, masses etc.).

In our previous paper \citep{JCAP_ET} we discussed the prospects of observing gravitationally lensed inspiral NS-NS events in the Einstein telescope.
As we emphasized there, this is important since the detection of gravitationally lensed source in GW detectors would be, for example, an invaluable
source of information concerning cosmography, complementary to standard ones independent of
the local cosmic distance ladder calibrations. Similar forecasts for the LISA mission were made by \citet{Sereno1,Sereno2}.

Here we update our previous study in two important aspects. First, we consider the whole class of DCOs (not only NS-NS). Second, instead of using point values of intrinsic merger rates of GW sources and a simplified phenomenological model of their evolution, as in \cite{JCAP_ET} we are using the results obtained by \citet{BelczynskiII} with stellar population synthesis evolutionary code {\tt StarTrack}.

The paper is organized as follows. In Section~\ref{sec:methodology}, we briefly recapitulate our methodology (referring to \citet{JCAP_ET} for detailed calculations) and review DCO catalog build from evolutionary population synthesis code which we use thereafter.
 Then, in Section~\ref{sec:rates} we present merger rates, forecasted yearly detection rates and forecasted lensing rates. Section~\ref{sec:conclusions}
contains the discussion of our results.

Throughout the paper we will assume flat FRW cosmological model as the one most supported by observations. In contrast to our previous paper where we assumed quintessential cosmological model with Chevalier-Polarski-Linder equation of state, here we confined ourselves to the $\Lambda$CDM model. This is exactly the same model as used by \citet{BelczynskiII} so we adopt it for consistency. In particular the expansion rate in this model reads:
\begin{equation} \label{H}
H(z) = H_0 \sqrt{\Omega_m (1+z)^3 + (1 - \Omega_m) }
\end{equation}
with $H_0 = 70\;km/s/Mpc$, $\Omega_m = 0.3$ as in \citet{BelczynskiII}. We will also adopt the notation: $E(z)=H(z)/H_0$ and ${\tilde r} = \int_0^{z} \frac{dz'}{E(z')}$ --- the non-dimensional comoving distance.


\section{Methodology} \label{sec:methodology}

As we already mentioned, we proceed in the same way as in our previous paper, so for more detailed calculations we refer the reader to \citep{JCAP_ET}. Here we just recap main points for clarity.
Our goal is to investigate the rate at which ET can see gravitationally lensed DCO sources. Therefore we need first to estimate the yearly detection rates of unlensed sources ${\dot N}(>\rho_0|z_s)$, then to assess the optical depth for lensing $\tau$ and finally combine these two to obtain the final prediction.
The yearly detection rate of DCO sources originating at redshift $z_s$ and producing the signal with S/N ratio exceeding the detector's threshold $\rho_0$
(we assume $\rho_0=8.$) can be expressed as:
\begin{equation} \label{Ndot}
{\dot N}(>\rho_0|z_s) = \int_0^{z_s} \frac{d {\dot N}(>\rho_0)}{dz} dz
\end{equation}
where
\begin{equation} \label{rate_nl}
\frac{d {\dot N}(>\rho_0)}{dz_s} = 4\pi \left( \frac{c}{H_0} \right)^3 \frac{\dot n_0(z_s)}{1+z_s} \;  \frac{{\tilde r}^2(z_s)}{E(z_s)} \; C_{\Theta}(x(z_s))
\end{equation}
is the rate at which we observe the inspiral DCO events (sources) that originate in the redshift
interval $[z, \; z + dz]$. The $C_{\Theta}(x(z_s))$ function captures (in a simplified way) the detector's performance and in particular it depends on
the distance parameters assumed here (after \citet{TaylorGair}) to be $r_0 = 1527 \; Mpc$ for the initial configuration and $r_0 = 1918 \; Mpc$ for the ``xylophone'' configuration. In this paper, unlike the previous one, we do not model the source evolution (i.e. the redshift dependence of intrinsic inspiral rate ${\dot n_0(z_s)}$) by an analytical formula but instead use the values of inspiral rates reported by \citet{BelczynskiII} for each redshift slice they considered.
The paper of \citet{BelczynskiII} contains the results of detailed population synthesis calculations performed with {\tt StarTrack} evolutionary code \citep{StarTrack}. Making (very thoroughly motivated) assumptions about star formation rate, galaxy mass distribution, stellar populations, their metallicities and galaxy metallicity evolution with redshift (``low-end'' and ``high-end'' cases), they evolved binary systems from ZAMS until the compact binary formation (after supernova explosions).
Because the compact object formation depends on the physics of common envelope (CE) phase of evolution and on SN explosion mechanism and both of them are to some degree uncertain, \citet{BelczynskiII} considered four scenarios: standard one and three of its modifications --- Optimistic Common Envelope (OCE), delayed SN explosion and high BH kicks scenario.
Common envelope phase is a crucial phase of binary evolution from the point of view of DCO formation. Namely, if the donor is on the Main Sequence (MS) or passes through the Hertzsprung Gap (HG) orbital energy is transferred to the whole star not to the envelope. This makes the envelope ejection difficult and such systems will merge preventing DCO formation. Physical reason for this is that MS stars have no clear distinction (entropy jump) between core and envelope. While this is also most probably the case of HG stars (as assumed in the ``standard scenario'') , yet it is also not clear at which evolutionary phase core-envelope structure actually develops. Hence the OCE scenario assumes that binaries with HG donor stars will lead to DCO formation. Standard scenario makes use of the so called ``rapid'' convection driven neutrino enhanced SN explosion mechanism. If this convection driven neutrino enhanced engine originates from the standing accretion shock instability, the explosion is delayed --- this is the ``delayed SN'' scenario. Its main effect is at the level of DCO mass spectrum, not so much affecting the inspiral rates. The last issue is that of the natal kicks, which are observationally supported in the case of NS. Natal kicks could be able to disrupt the binary preventing DCO formation. Standard scenario makes a conservative assumption that the velocity of the natal kick is reduced by some factor related to possible fallback of some amount of matter to the newborn compact object. ``High BH kicks'' scenario assumes that newborn compact object receives the full kick. For more details, see \citet{BelczynskiII} and references therein.

\begin{figure}
\begin{center}
\includegraphics[angle=270,width=70mm]{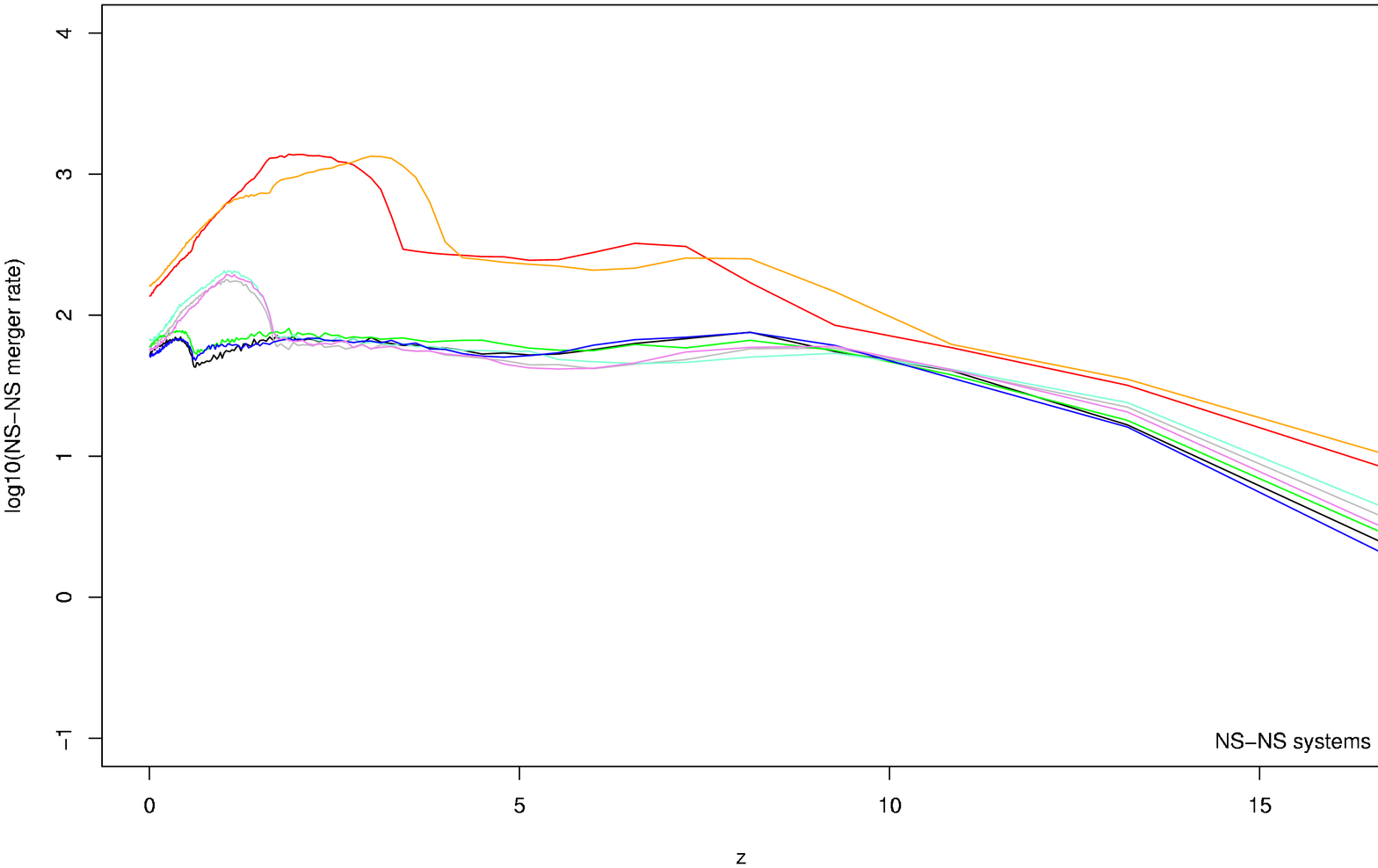}
\includegraphics[angle=270,width=70mm]{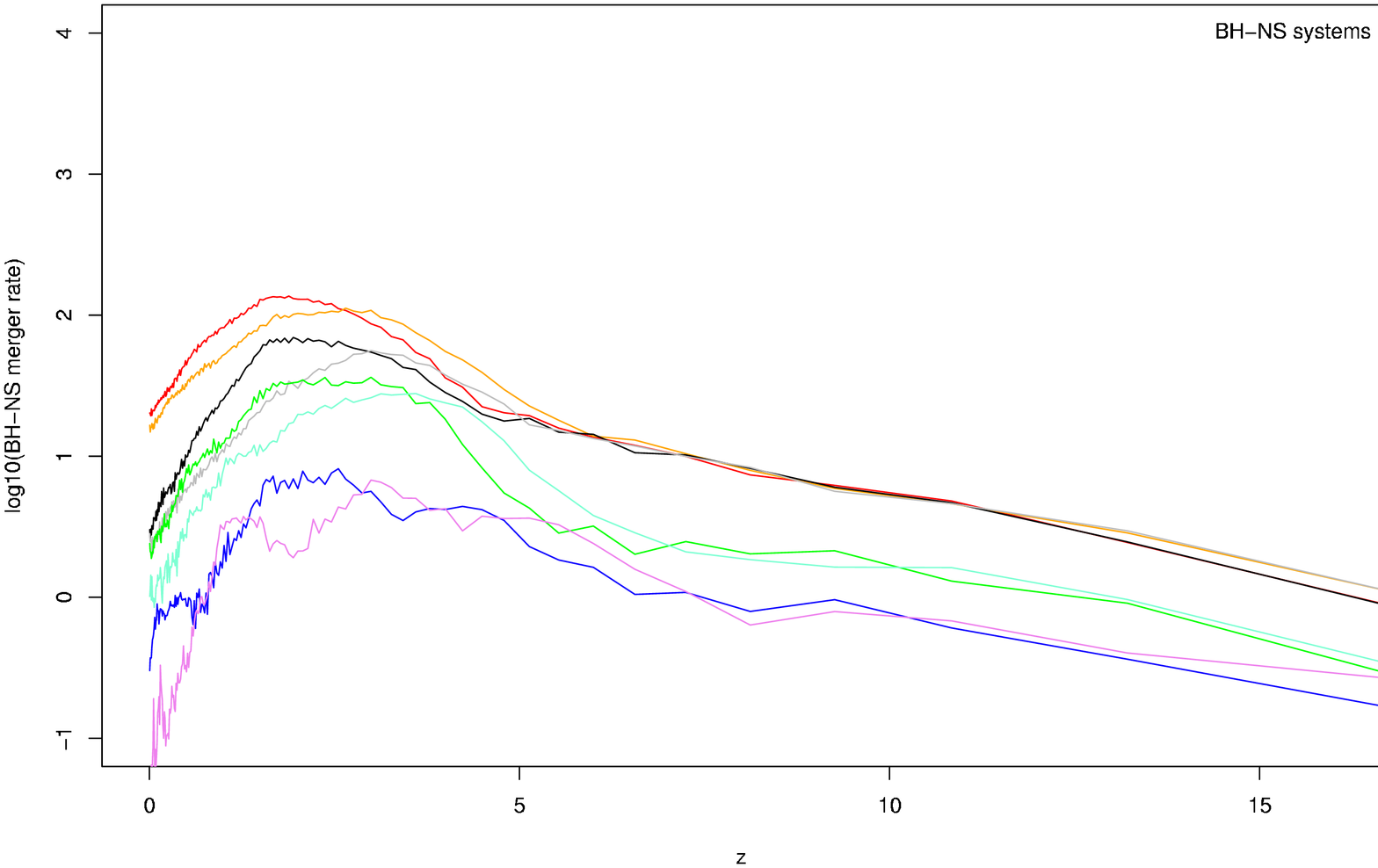}
\includegraphics[angle=270,width=70mm]{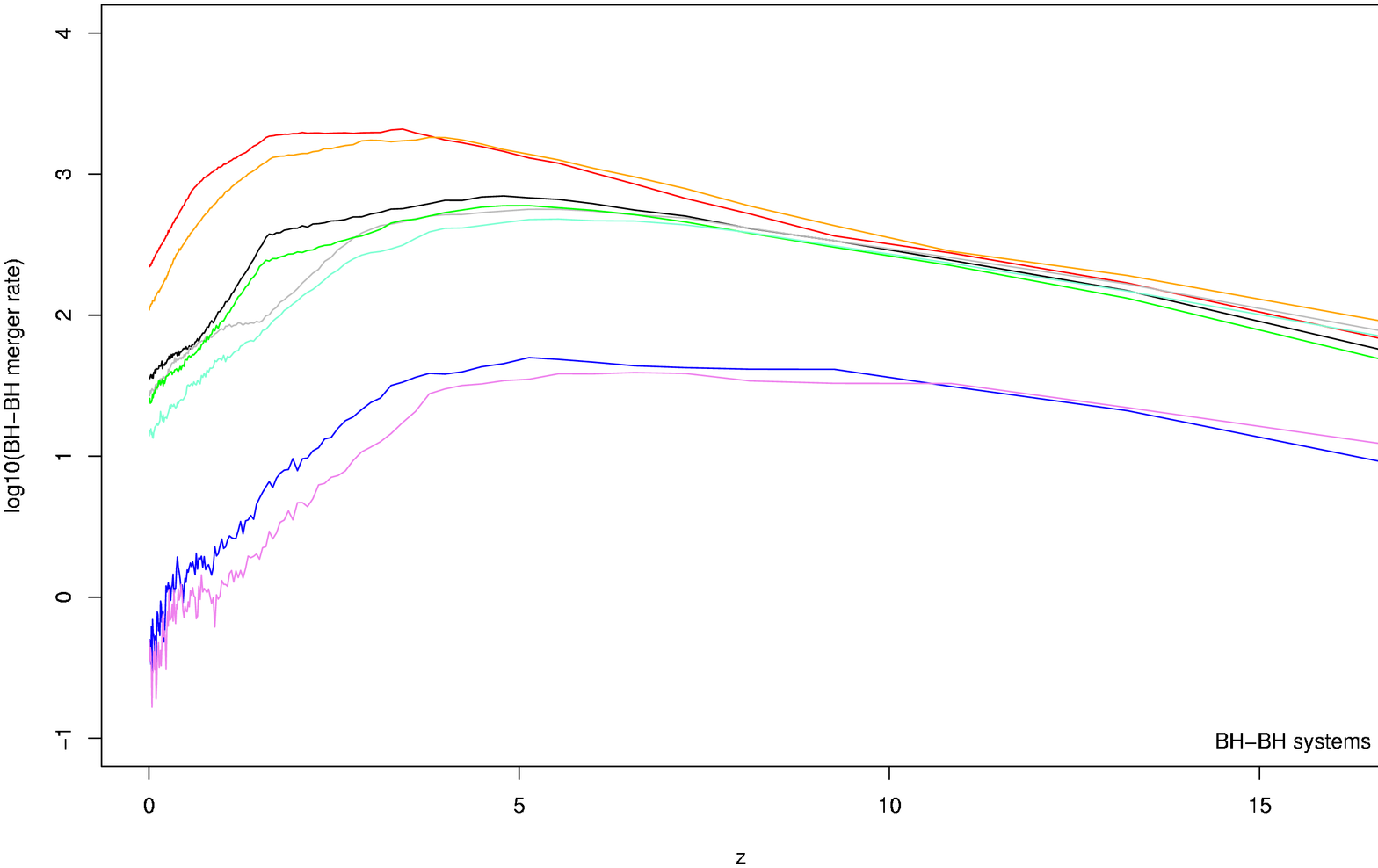}
\includegraphics[angle=270,width=40mm]{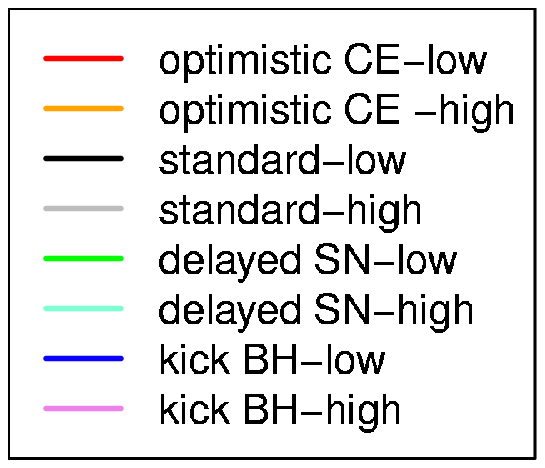}
\end{center}
\caption{Intrinsic merger rates in all possible scenarios according to simulations of \citet{BelczynskiII}.
\label{merger_rate}}
\end{figure}

We consider all these scenarios retaining their names as in \citep{BelczynskiII}. We have taken the data from the website  {\tt http:www.syntheticuniverse.org}, more specifically the so called ``rest frame rates'' in cosmological scenario.
Fig~\ref{merger_rate} shows the intrinsic merger rates ${\dot n_0}(z)$ according to the data from \citet{BelczynskiII}.

We have assumed the following values of the chirp masses: $1.2\;M_{\odot}$ for NS-NS, $3.2\;M_{\odot}$ for BH-NS and $6.7\;M_{\odot}$ for BH-BH systems.
According to \citet{BelczynskiI}, these values represent average chirp mass for each category of DCO simulated by population synthesis.

Concerning gravitational lensing we adopt the same approach as in our previous paper \citep{JCAP_ET}, i.e. we assume conservatively that the population of lenses comprise only elliptical galaxies. Spiral galaxies can also act as lenses, but ellipticals are more massive and indeed they dominate in all strong lensing surveys.
Consequently, we will model the lenses as singular isothermal spheres (SIS) which is a good approximation of early type galaxies \citep{Koopmans09}.

From the physical point of view, mass (and its distribution) is the most important parameter for gravitational lensing. On the other hand, it is the luminosity function $\varphi(L)$ that has become the standard way to characterize populations of galaxies. However, there is now a noticeable trend to use (stellar) mass, size and velocity distributions as more informative and less biased characteristics. The most recent advances in this field can be found in very important papers by \citet{Bernardi2010, Bernardi2013}. In particular, they noticed that besides large systematic uncertainties concerning stellar M/L ratio the massive end of the stellar mass function is sensitive to which light profile one fits to the most luminous galaxies. Fortunately, the SIS model assumed here depends on stellar velocity dispersion $\sigma$ and the velocity dispersion distribution of early type galaxies has been rather well modeled by \citet{Sheth2003}, then modified by \citet{Mitchell2005}. More recently \citet{Choi2007} using a sample from SDSS Data Release 5, derived the velocity dispersion function which since then became a standard in the studies of gravitational lensing statistics.
Consequently, we model the velocity dispersion distribution in the population of lensing galaxies as a modified Schechter function $\frac{d n}{d \sigma} = n_{*} \left( \frac{\sigma}{\sigma_{*}} \right)^{\alpha} \exp{\left( - \left( \frac{\sigma}{\sigma_{*}} \right)^{\beta} \right)} \frac{\beta}{\Gamma (\frac{\alpha}{\beta}) } \frac{1}{\sigma}$. And for the parameters $n_{*}$,$\sigma_{*}$,${\alpha}$ and $\beta$ we take the values obtained by \citet{Choi2007} . Let us note that \citet{Bernardi2010} also provided the velocity distribution fits on even better data (SDSS DR6 which was corrected for low- velocity dispersion bias present in DR5). However, their velocity function is for galaxies of all types hence we used the values from \citep{Choi2007}. Using the fixed values for parameters in the velocity distribution function also deserves a comment. Namely, the evolution of velocity dispersion and number density of galaxies (i.e. the redshift dependence of $\sigma_{*}$ and $n_{*}$ parameters) can introduce considerable uncertainties and bias the optical depth for lensing. However, the works investigating this issue \citep{Mitchell2005,Natarayan2007,Oguri2012} have concluded that galaxy evolution does not significantly affect lensing statistics.
Clearly, our assumptions concerning velocity distribution function of galaxies could be improved in many aspects, but these subtle issues are likely to be subdominant comparing with the properties of the population of sources (like DCO formation scenarios). We will return to this discussion in the final section.


The SIS lens produces two images separated by $\theta_E$ (the Einstein radius in radians), one brighter and one fainter w.r.t. unlensed source,  from which the signal comes with a relative time delay $\Delta t := \Delta t_0(\sigma;{\tilde r_l},{\tilde r_s}) y$ where $y$ is nondimensional (i.e. in units of the Einstein radius) misalignment angle between source and lens. Exact formulae are well known, but see also our previous paper \citep{JCAP_ET} whose notation we comply with. In order to say we have seen a lensed GW source the fainter image should have S/N ratio higher than threshold, so the misalignment must not exceed some limiting value $y_{max}$. This influences the elementary cross-section: $S_{cr} = \pi \theta_E^2 y_{max}^2$ and further propagates into formulae for optical depth for lensing. In particular differential (with respect to the lens redshift) optical depth for lensing reads:
\begin{equation} \label{dtaudz}
\frac{d \tau}{d z_l} = 16 \pi^3 \left( \frac{c}{H_0} \right)^3 \frac{{\tilde r}^2_{ls} {\tilde r}^2_l}{{\tilde r}^2_s E(z_l)} y_{max}^2 n_{*} \left( \frac{\sigma_{*}}{c} \right)^4 \frac{\Gamma \left( \frac{4 + \alpha}{\beta} \right) }{\Gamma \left( \frac{\alpha}{\beta} \right)}
\end{equation}
and the total optical depth is:
\begin{equation} \label{tau}
\tau = \frac{16}{30} \pi^3 \left( \frac{c}{H_0} \right)^3 {\tilde r}_s^3 \left( \frac{\sigma_{*}}{c} \right)^4 n_{*} \frac{\Gamma \left( \frac{4 + \alpha}{\beta} \right) }{\Gamma \left( \frac{\alpha}{\beta} \right)} y_{max}^2
\end{equation}

The above formulae are valid only in the case of a continuous search. If instead the survey has a finite duration $T_{surv}$ some of the events (whose signals come near the beginning or the end of the survey) would be lost because of lensing time delay --- i.e. we would register the signal from just one image and cannot tell that in fact the event was lensed. Such case can be handled by a proper correction to (\ref{dtaudz}) or (\ref{tau}), e.g.:
\begin{equation} \label{tauDT}
\tau_{\Delta t} =  \tau \left[ 1 - \frac{1}{7} \frac{\Gamma \left( \frac{\alpha+8}{\beta} \right)}{\Gamma \left( \frac{\alpha+4}{\beta} \right)} \frac{{\Delta t}_{*}}{T_{surv}} \right]
\end{equation}
where: $\Delta t_{*} = \frac{32 \pi^2}{H_0} {\tilde r}_s \left( \frac{\sigma_{*}}{c} \right)^4 y_{max} $ (for detailed calculations see \citet{JCAP_ET}).

These ingredients can be combined to determine cumulative yearly detection of lensed events up to the source redshift $z_s$:
\begin{equation} \label{lensing_rate}
{\dot N}_{lensed}(z_s) = \int_0^{z_s} \tau(z_s, y_{max}, T_{surv}) \frac{d {\dot N}(>\rho_0)}{dz} dz
\end{equation}

\section{Detection rates and gravitational lensing statistics} \label{sec:rates}

\begin{figure}
\begin{center}
\includegraphics[angle=270,width=70mm]{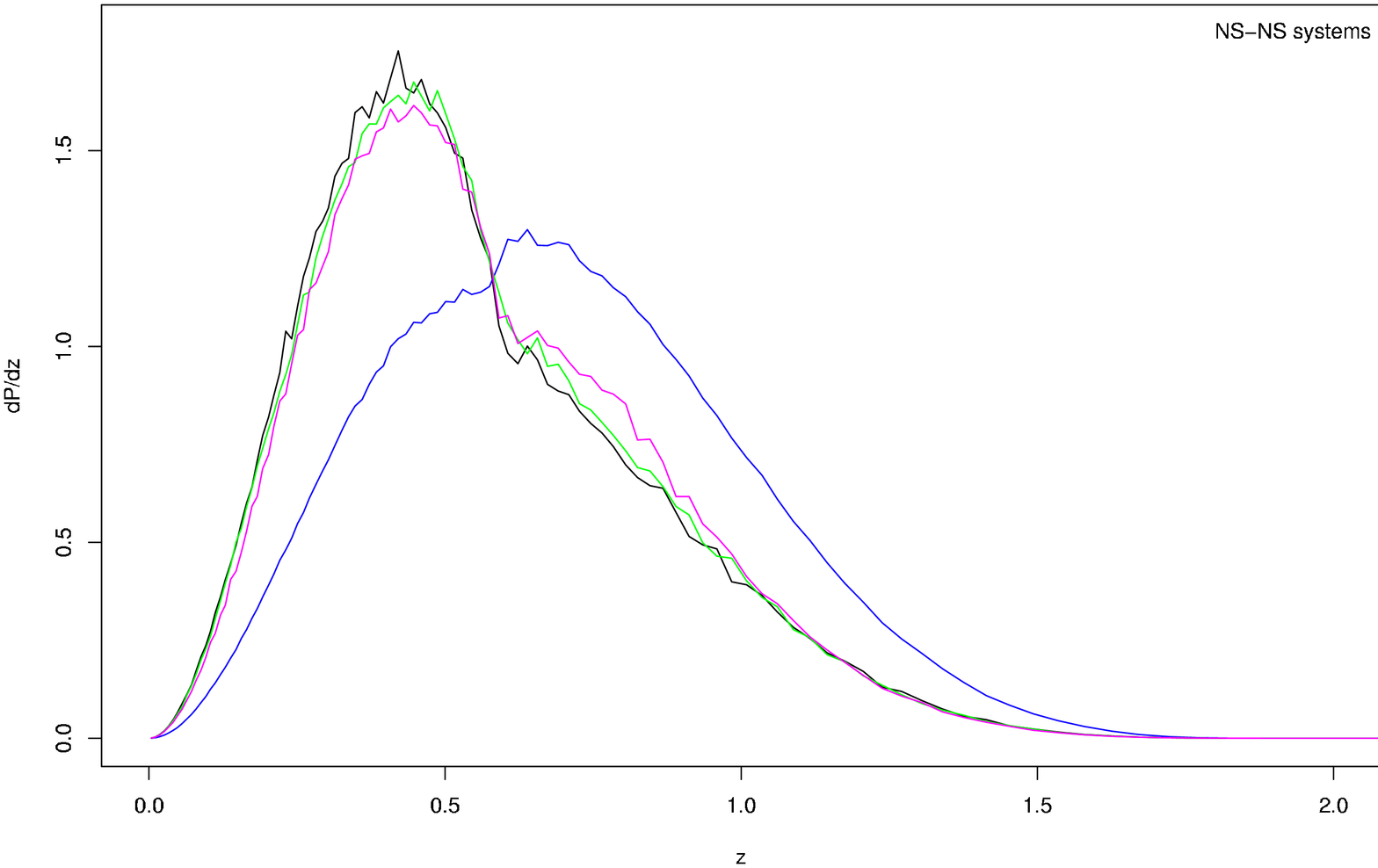}
\includegraphics[angle=270,width=70mm]{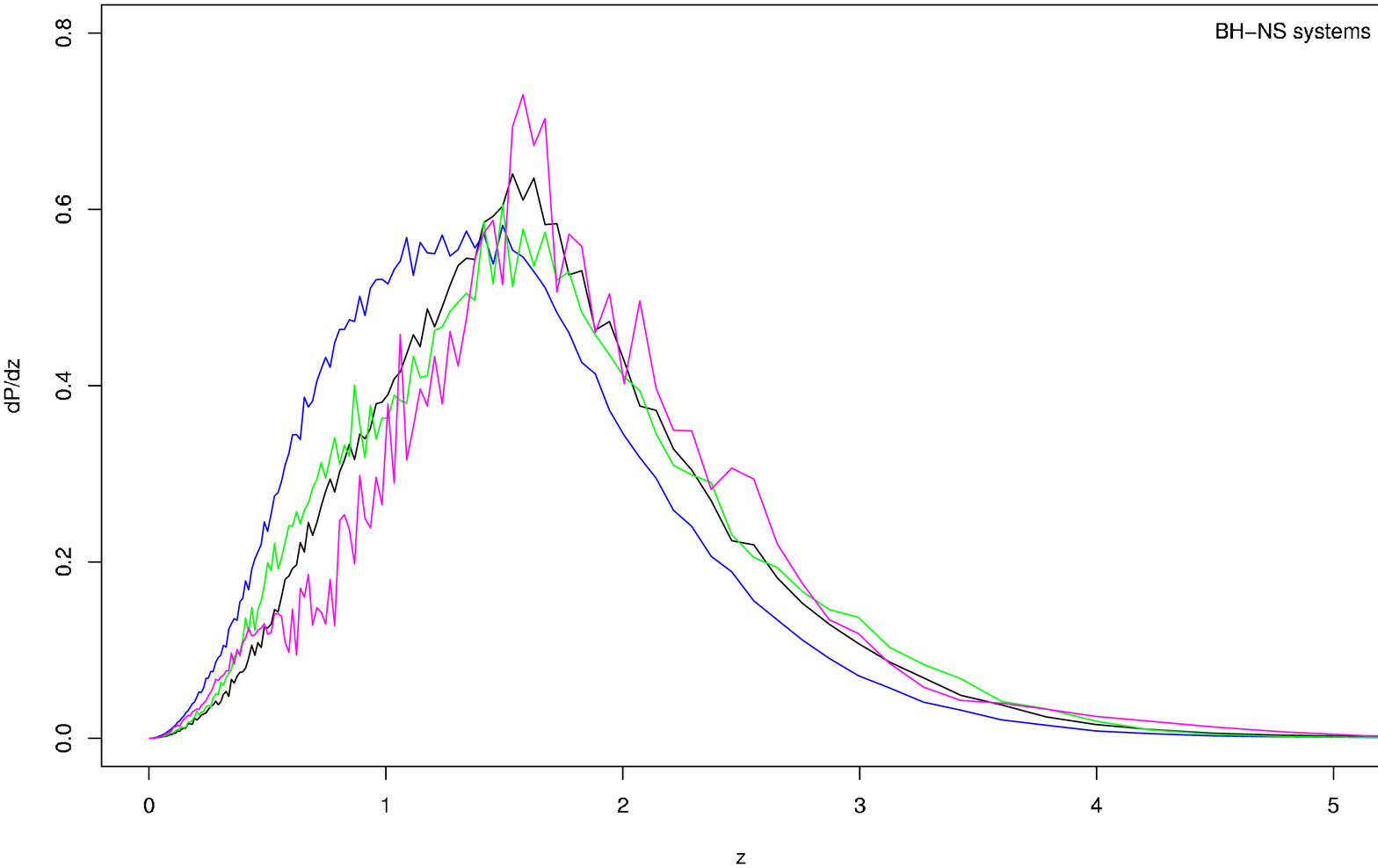}
\includegraphics[angle=270,width=70mm]{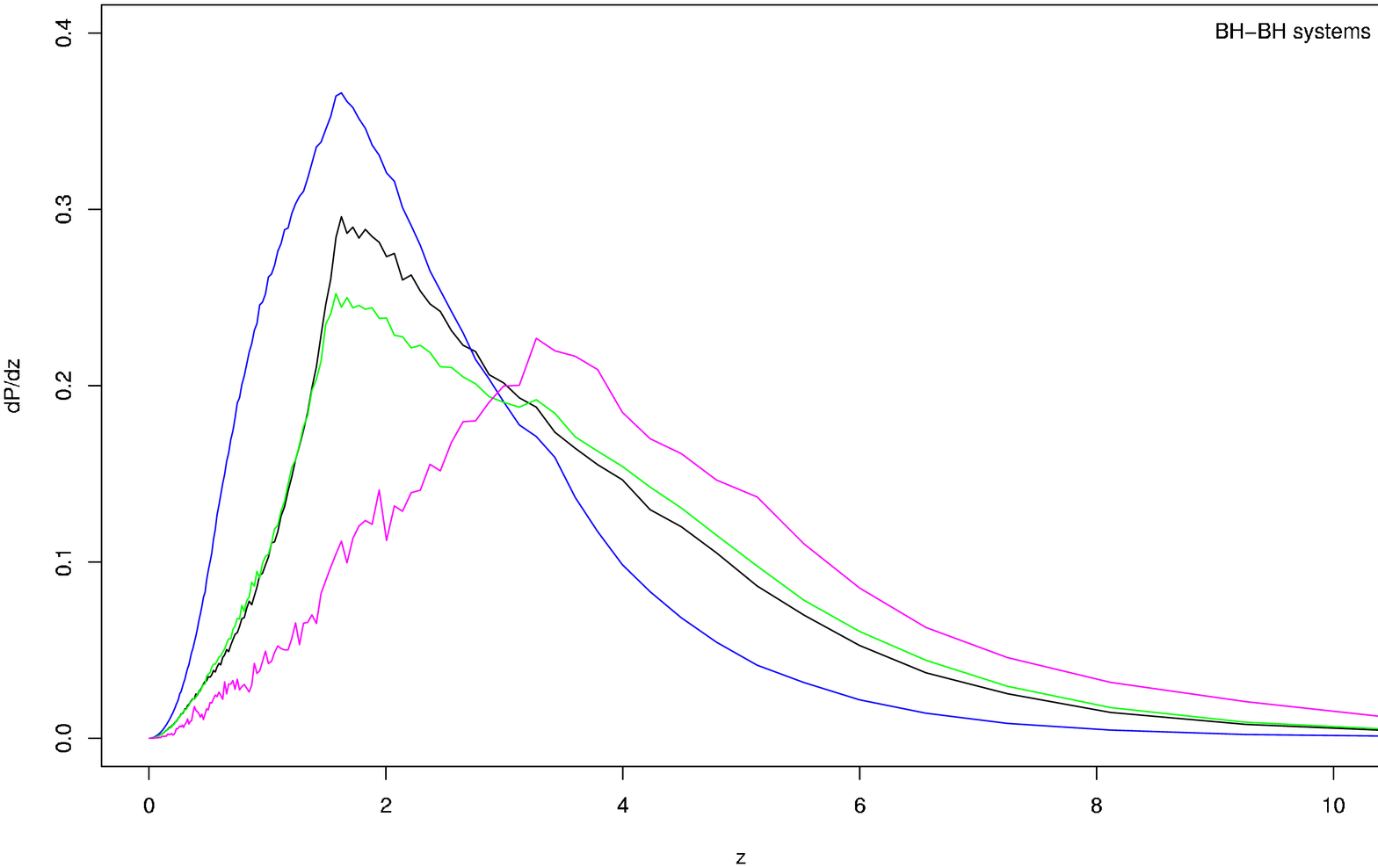}
\includegraphics[angle=270,width=70mm]{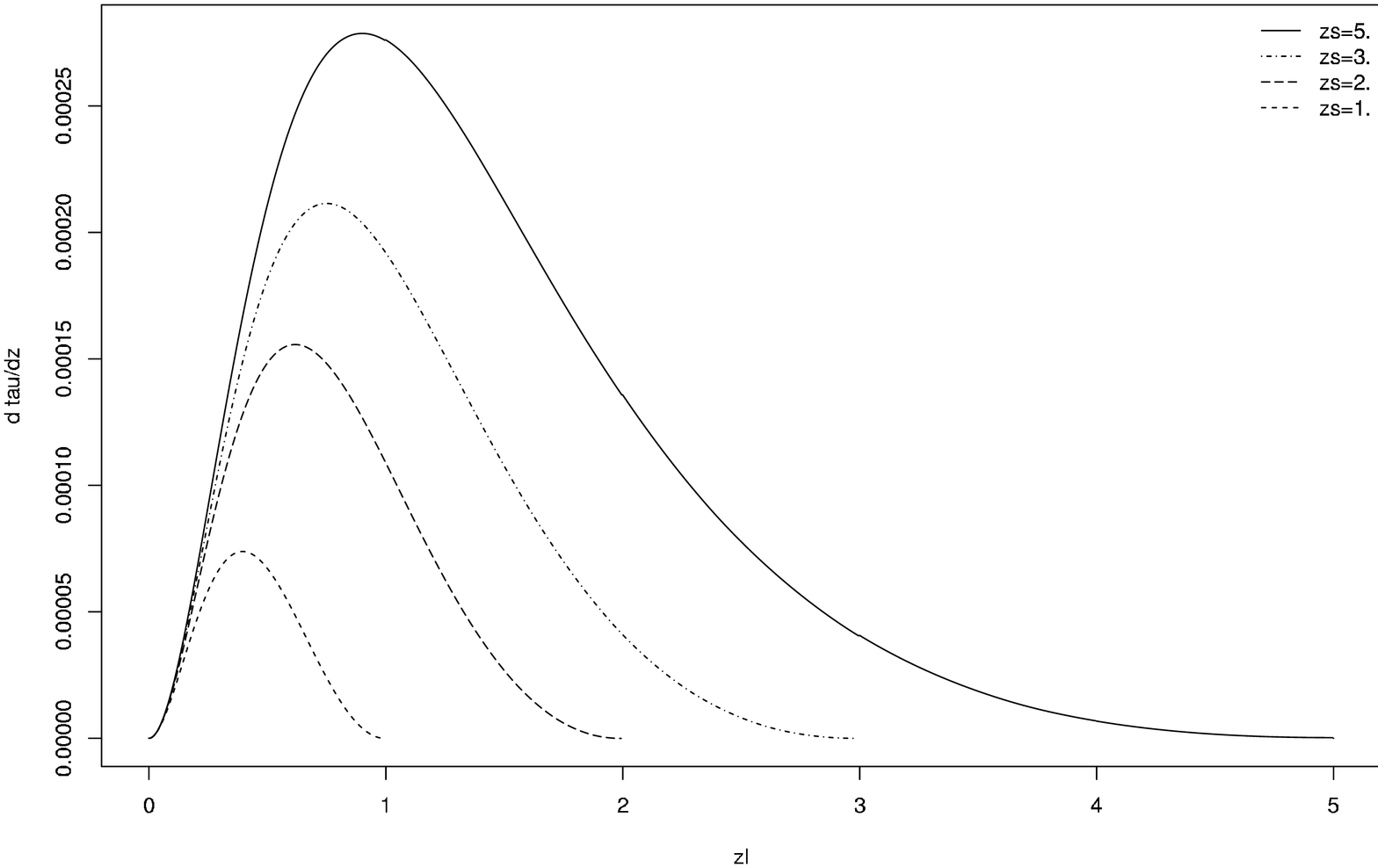}
\end{center}
\caption{Probability density of DCO inspiral events as a function of redshift for the initial ET design. Different colors refer to different scenarios:
black -- standard, blue -- OCE, green -- delayed SN, magenta -- high BH kicks. Low-end metallicity evolution was assumed. Lower right panel shows
differential optical depth for lensing as a function of lens redshift for different source redshifts.
\label{diff_rate}}
\end{figure}

Here we present the results obtained according to the methodology outlined in Section~\ref{sec:methodology}. Figure~\ref{diff_rate} shows combined probability density of NS-NS, BH-NS, BH-BH inspiral rates. Different colors correspond to different DCO evolutionary scenarios as explained in the Figure caption. Differential inspiral event rates (\ref{rate_nl}) have been normalized to respective probability distribution functions which are better suited for comparisons. Differential optical depth for lensing as a function of $z_l$ is also shown for sources located at redshifts from $z_s=1$ to $z_s=5$. On Figure~\ref{diff_rate} only one particular metallicity evolution scenario ("low-end") was shown. Moreover the initial ET configuration was assumed. We adopted such philosophy  concerning figures in order not to proliferate them and maintain their transparency. This is especially relevant to figures plotted in the logarithmic scale where differences between scenarios would hardly be visible. On the other hand data in Tables are comprehensive.

\begin{figure}
\begin{center}
\includegraphics[angle=270,width=70mm]{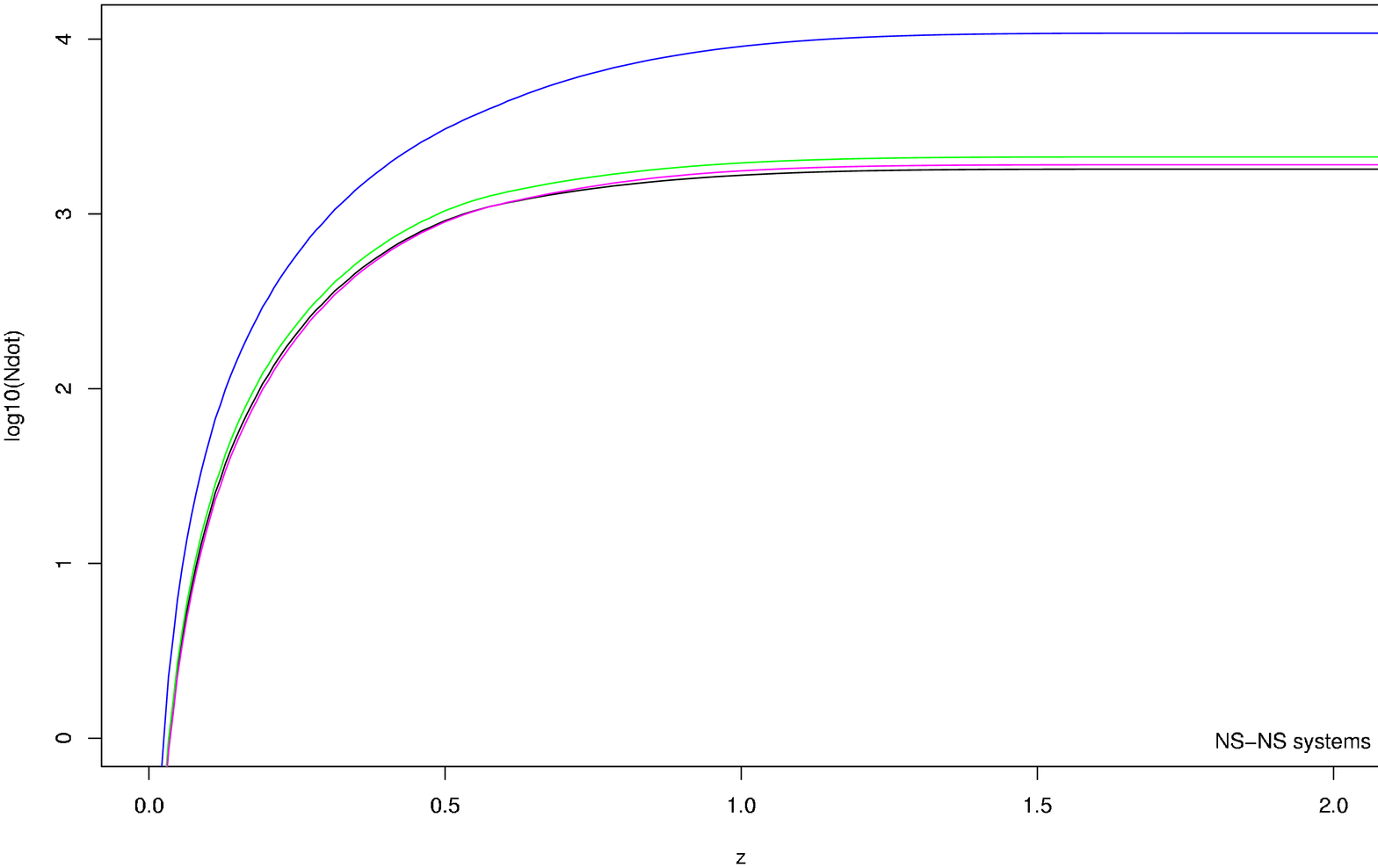}
\includegraphics[angle=270,width=70mm]{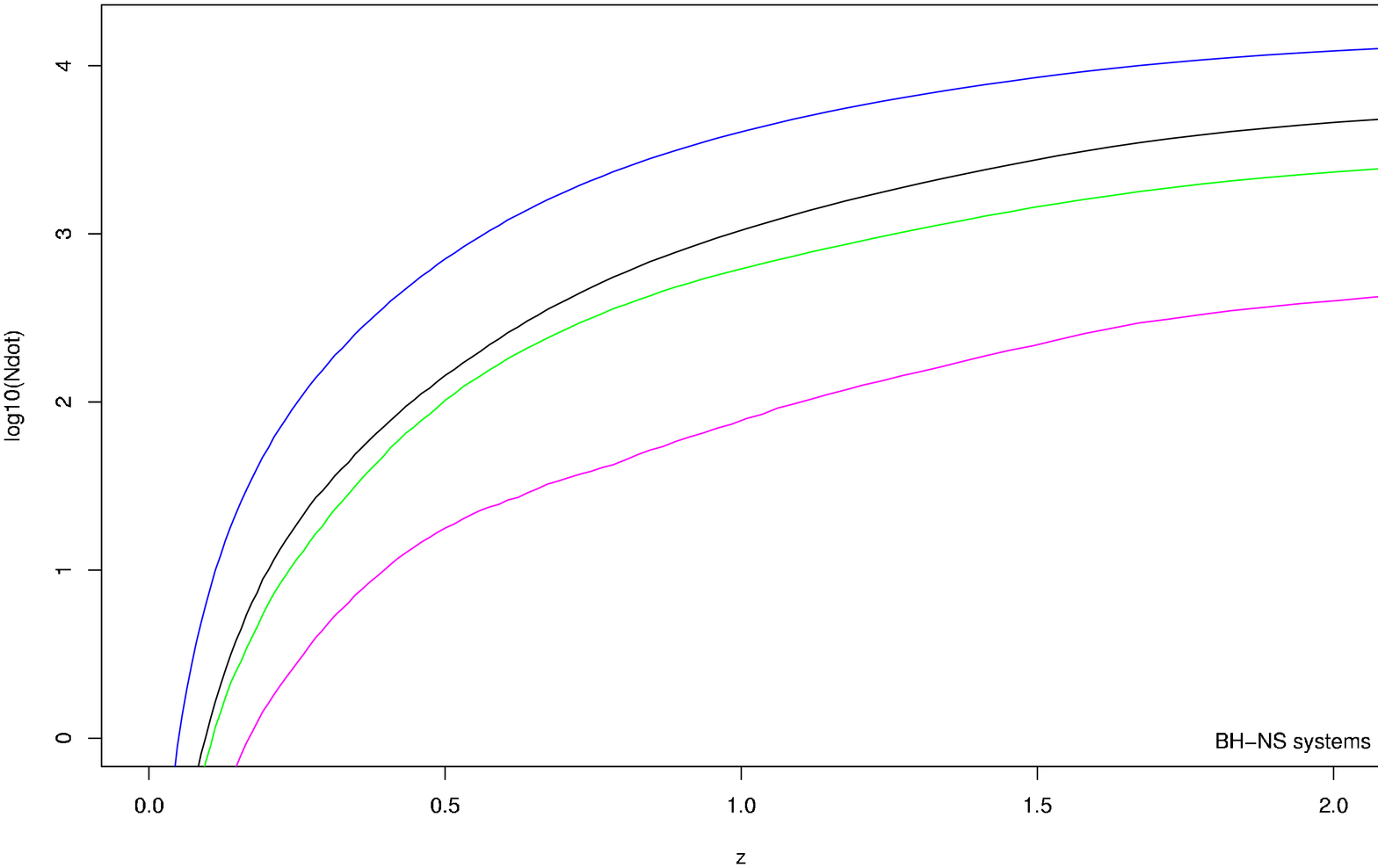}
\includegraphics[angle=270,width=70mm]{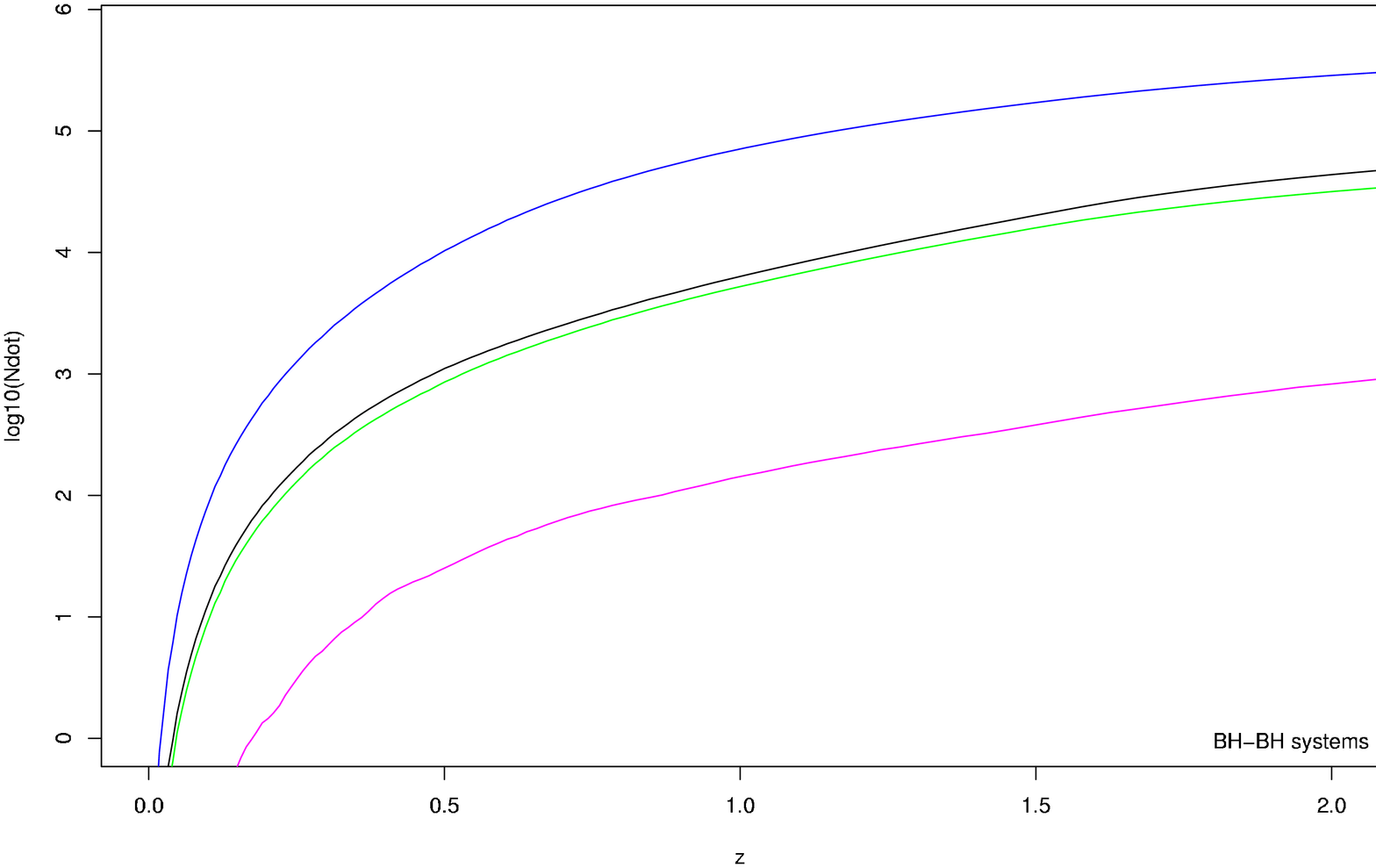}

\end{center}
\caption{Yearly detection rate (as a function of source redshift) of DCO inspiralling binaries for different evolutionary scenarios. Low-end metallicity evolution scenario assumed. Prediction is for the Einstein Telescope in its initial configuration. Logarithmic scale (base 10) is adopted.
\label{rates}}
\end{figure}

\begin{figure}
\begin{center}
\includegraphics[angle=0,width=70mm]{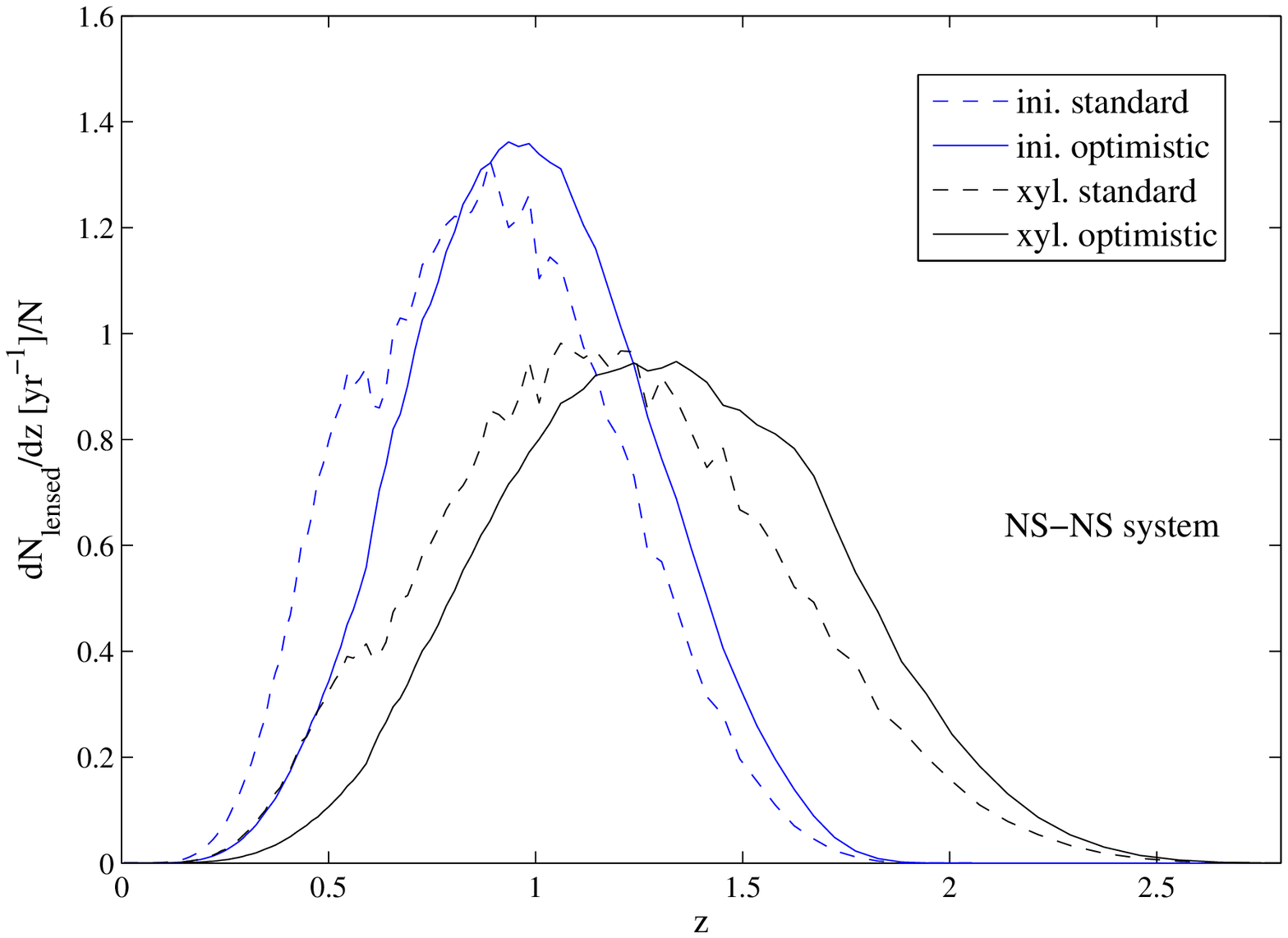}
\includegraphics[angle=0,width=70mm]{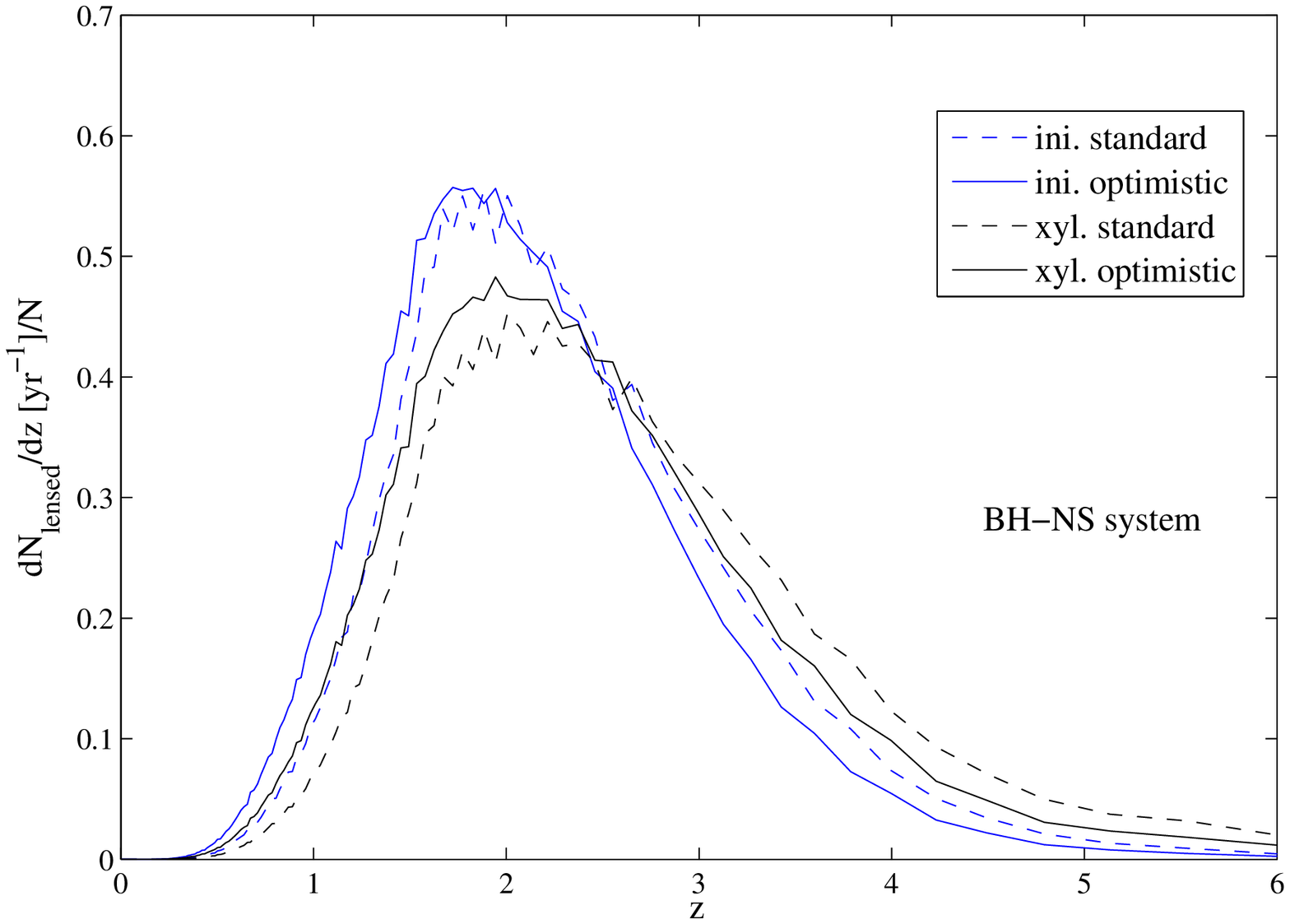}
\includegraphics[angle=0,width=70mm]{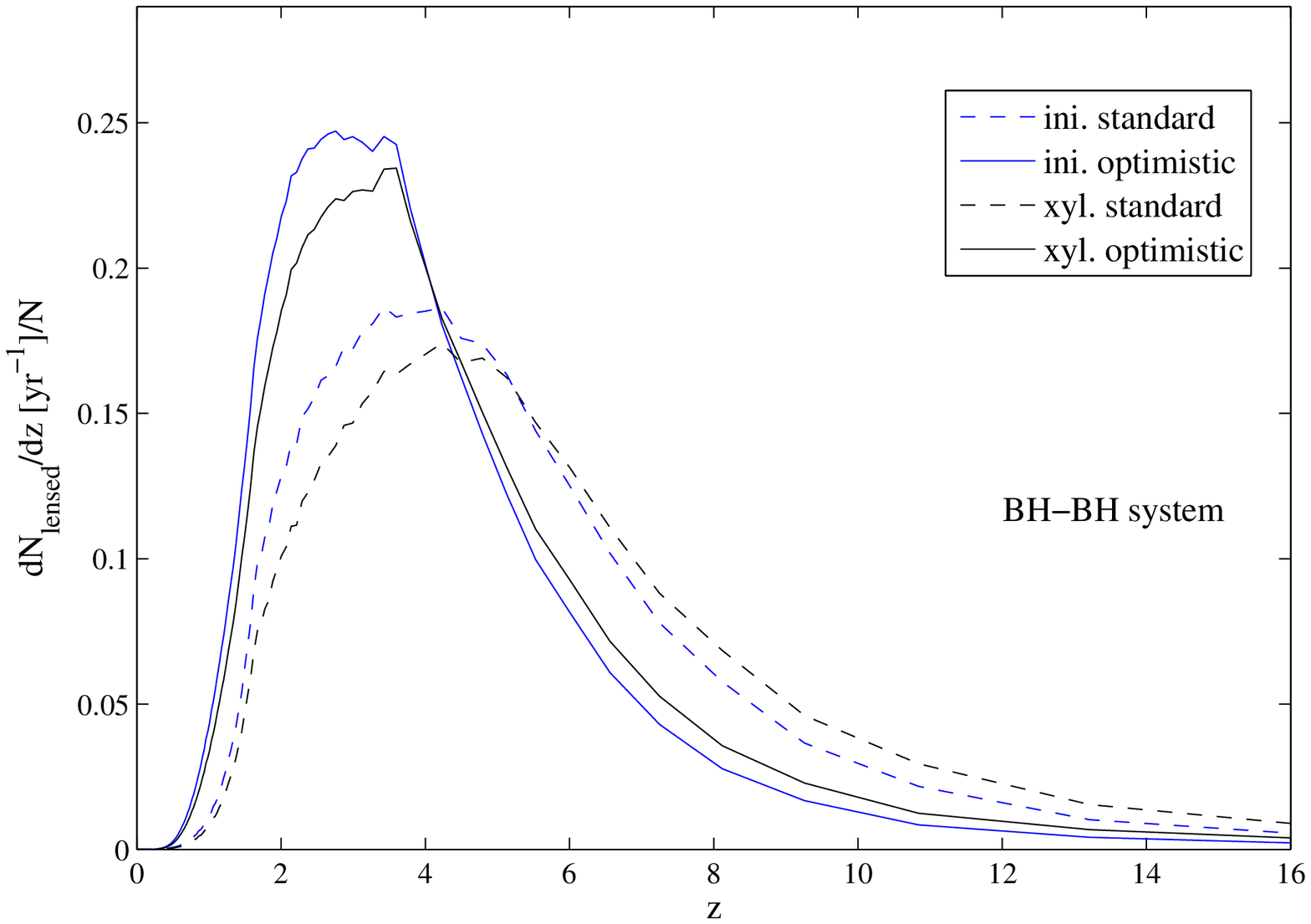}
\end{center}
\caption{Differential lensing rate $ \frac{1}{{\dot N}_{lensed}} \frac{d{{\dot N}_{lensed}}}{dz}$(as a function of source redshift) of DCO inspiralling binaries for two different evolutionary scenarios. Black line is for the standard scenario, blue one --  for the optimistic CE.Low-end metallicity evolution scenario assumed. Predictions for the Einstein Telescope in its initial and "xylophone" configuration are shown. Solid line denotes initial configuration, long dashed one- ``xylophone'' configuration.
\label{diff lensing}}
\end{figure}

\begin{figure}
\begin{center}
\includegraphics[angle=270,width=70mm]{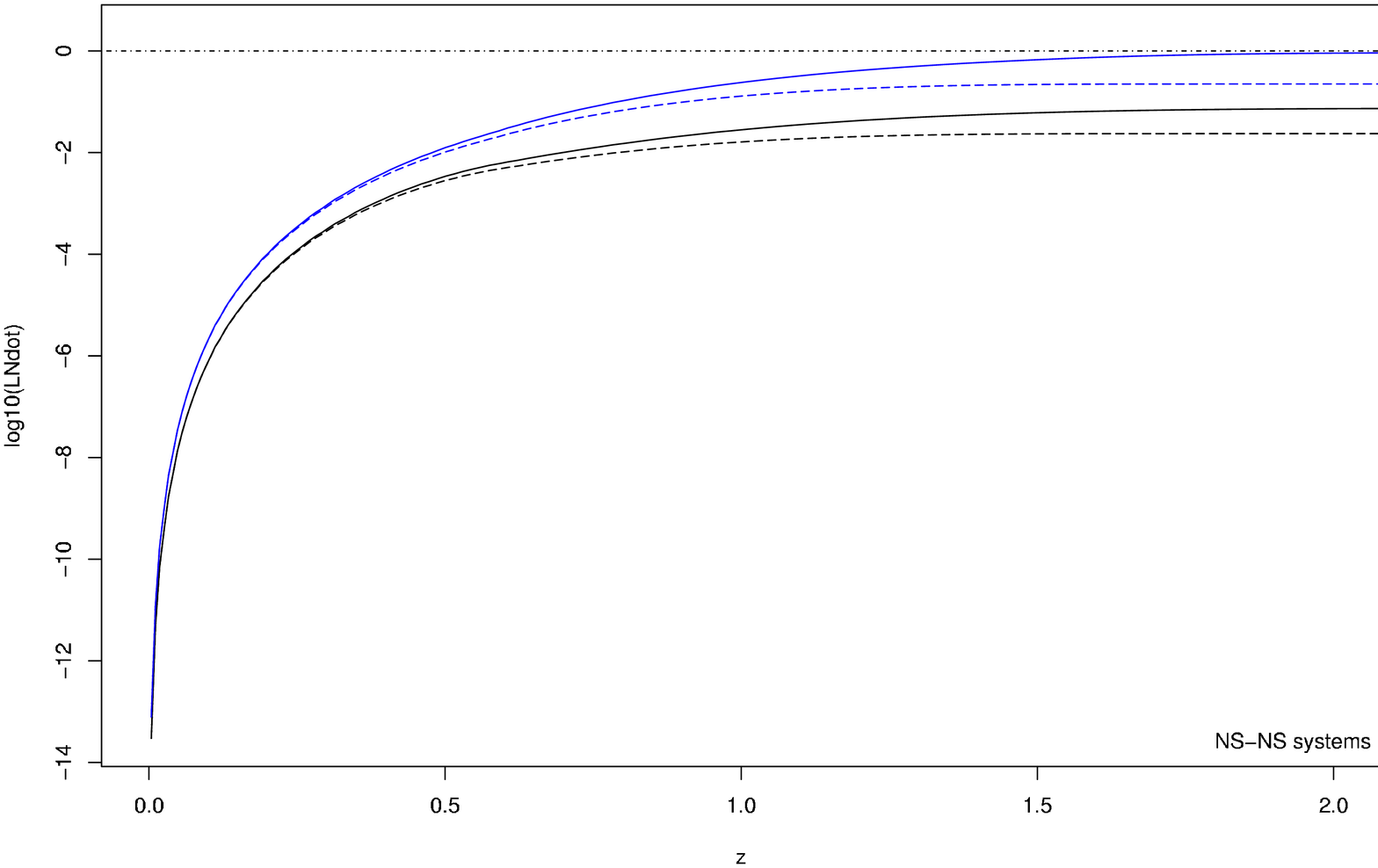}
\includegraphics[angle=270,width=70mm]{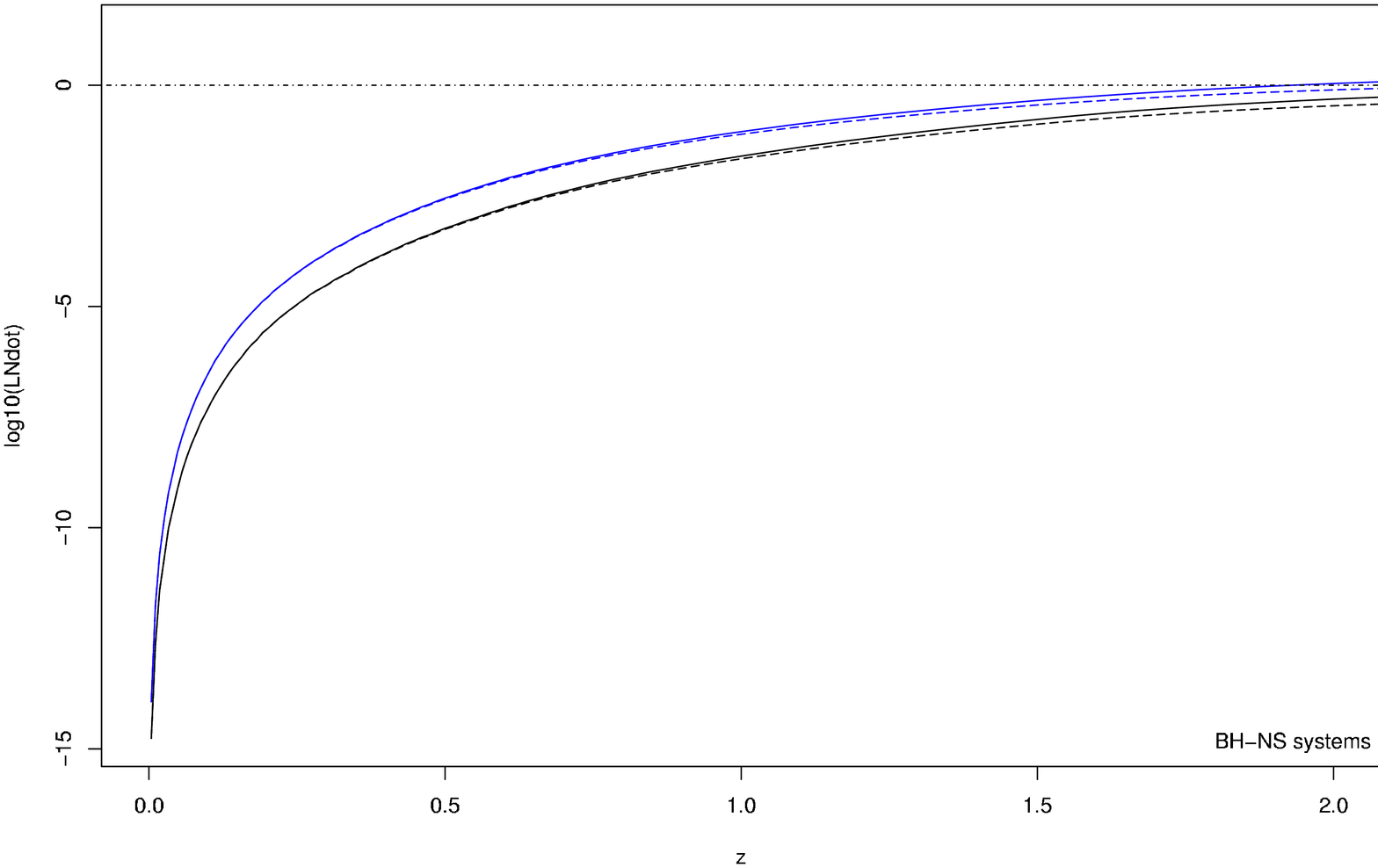}
\includegraphics[angle=270,width=70mm]{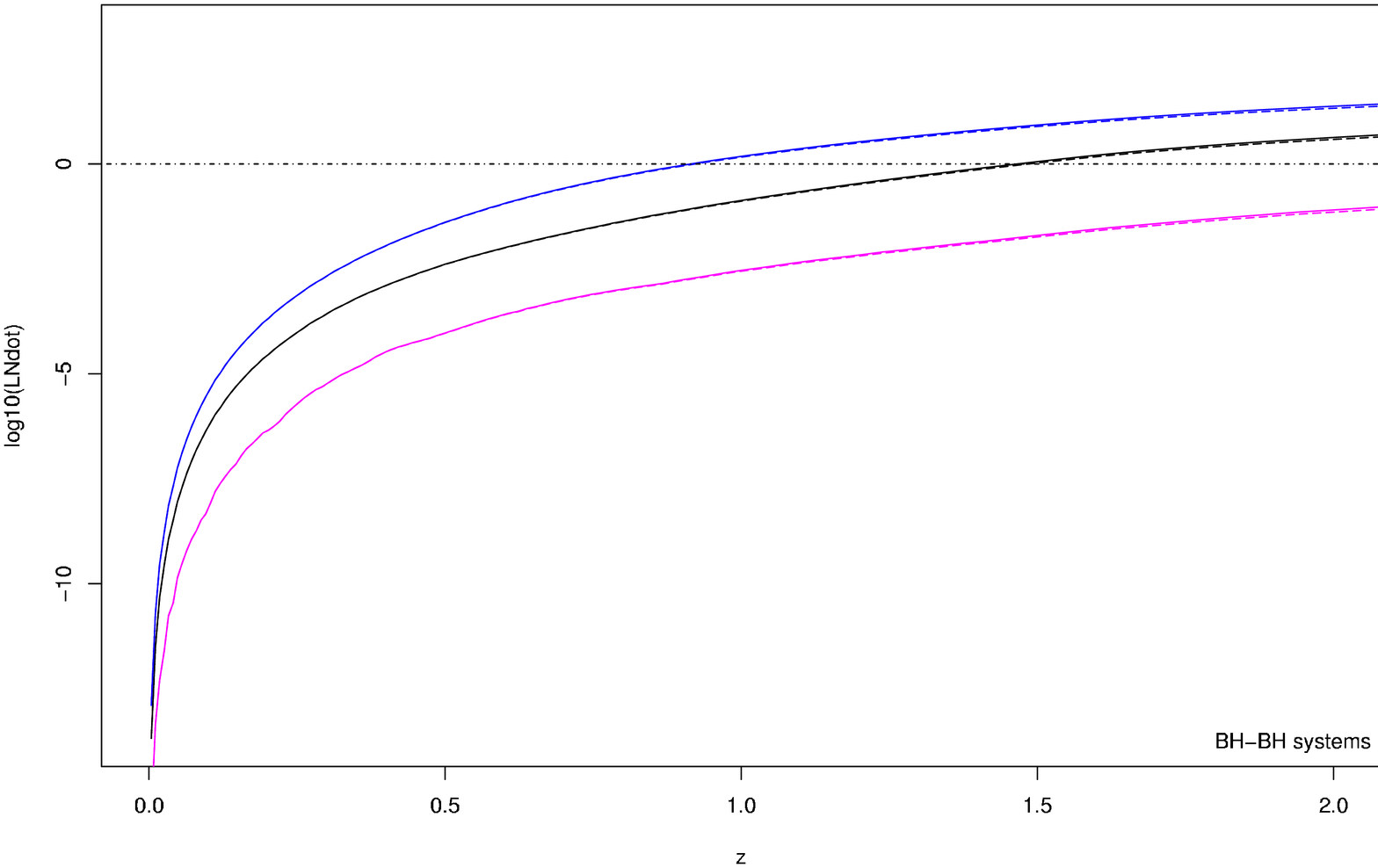}

\end{center}
\caption{Expected yearly detection rate (as a function of source redshift) of lensed DCO signals for different evolutionary scenarios. Low-end metallicity evolution scenario assumed. Prediction is for the Einstein Telescope in its initial configuration. Logarithmic scale (base 10) is adopted.
\label{lensing rates}}
\end{figure}

\begin{table*}[ht]
\caption{Yearly detection rate of inspiralling DCOs of different classes under different evolutionary scenarios. ``High-end'' metallicity evolution assumed.
Predictions for the Einstein Telescope in the initial and ``xylophone'' configuration.}
\label{rates-hi}
\begin{center}  
\begin{tabular}{cccccc}

\hline \
Yearly detection rate & standard & optimistic CE & delayed SN & high BH kicks \\

\hline \\

${\dot N}(>\rho_0)$ [$yr^{-1}$] for NS-NS &&&&& \\
initial design & $3834.8$ & $11631.8$ & $4303.4$ & $3683.1$ \\
``xylophone'' design & $6948.8$ & $23399.3$ & $7832.8$ & $6879.7$ \\
\\
\hline \\

${\dot N}(>\rho_0)$ [$yr^{-1}$] for BH-NS &&&&& \\
initial design & $3468.1$ & $11539.3$ & $1939.9$ & $434.1$ \\
``xylophone'' design & $6004.5$ & $17704.5$ & $3302.5$ & $715.8$ \\
\\
\hline \\

${\dot N}(>\rho_0)$ [$yr^{-1}$] for BH-BH &&&&& \\
initial design & $105703.3$ & $494917.5$ & $81848.4$ & $4955.6$ \\
``xylophone'' design & $144801.9$ & $619555.5$ & $113879.8$ & $7411.1$ \\
\\
\hline \\

${\dot N}(>\rho_0)$ [$yr^{-1}$] for total &&&&& \\
initial design & $113006.2$ & $518088.5$ & $88091.7$ & $9072.8$ \\
``xylophone'' design & $157755.2$ & $660659.3$ & $125015.1$ & $15006.6$ \\
\\
\hline

\end{tabular}\\
\end{center}
\end{table*}

\begin{table*}[ht]
\caption{Yearly detection rate of inspiralling DCOs of different classes under different evolutionary scenarios. ``Low-end'' metallicity evolution assumed.
Predictions for the Einstein Telescope in the initial and ``xylophone'' configuration.} \label{rates-low}
\begin{center}  
\begin{tabular}{cccccc}

\hline \
Yearly detection rate & standard & optimistic CE & delayed SN & high BH kicks \\

\hline \\

${\dot N}(>\rho_0)$ [$yr^{-1}$] for NS-NS &&&&& \\
initial design & $1805.2$ & $10829.3$ & $2120.2$ & $1912.5$ \\
``xylophone'' design & $3016.4$ & $23817.4$ & $3543.3$ & $3218.3$ \\
\\
\hline \\

${\dot N}(>\rho_0)$ [$yr^{-1}$] for BH-NS &&&&& \\
initial design & $6510.9$ & $15622.6$ & $3379.6$ & $606.6$ \\
``xylophone'' design & $10178.1$ & $22730.4$ & $5297.2$ & $983.6$ \\
\\
\hline \\

${\dot N}(>\rho_0)$ [$yr^{-1}$] for BH-BH &&&&& \\
initial design & $158495.4$ & $633449.3$ & $123405.1$ & $7369.5$ \\
``xylophone'' design & $208499.4$ & $773055.9$ & $164615.1$ & $10588.3$ \\
\\
\hline \\

${\dot N}(>\rho_0)$ [$yr^{-1}$] for total &&&&& \\
initial design & $166811.5$ & $659901.2$ & $128904.9$ & $9888.5$ \\
``xylophone'' design & $221693.9$ & $819603.8$ & $173455.7$ & $14790.1$ \\
\\
\hline

\end{tabular}\\
\end{center}
\end{table*}

\begin{table*}[ht]
\footnotesize 
\caption{Expected numbers of lensed GW events from inspiralling DCOs of different classes under different evolutionary scenarios. ``High-end'' metallicity evolution assumed.
Predictions for the Einstein Telescope in the initial and ``xylophone'' configuration.}
\label{lensing-hi}
\begin{center}  
\begin{tabular}{cccccc}

\hline
ET configuration  &standard&optimistic&delayed SN& high BH kicks  \\
$T_{surv}$ &{\scriptsize (1yr; 5yrs;continuous)}&{\scriptsize (1yr; 5yrs;continuous)}
&{\scriptsize (1yr; 5yrs;continuous)}&{\scriptsize (1yr; 5yrs;continuous)}\\

\hline\\
NS-NS & & & & \\
initial design &(0.06; 0.07; 0.07)&(0.2; 0.2; 0.2)&(0.07; 0.07; 0.08)&(0.06; 0.06; 0.07)\\
xylophone &(0.2; 0.2; 0.2)&(0.7; 0.8; 0.8)&(0.2; 0.2; 0.2)&(0.2; 0.2; 0.2)\\
\\
\hline\\
BH-NS & & & & \\
initial design &(0.4; 0.5; 0.5)&(1.1; 1.3; 1.3)&(0.2; 0.3; 0.3)&(0.05; 0.05; 0.05)\\
xylophone &(0.9; 1.1; 1.1)& (2.1; 2.4; 2.5)&(0.5; 0.6; 0.6)&(0.1; 0.1; 0.1)\\
\\
\hline\\
BH-BH & & & & \\
initial design &(30.3; 36.1; 37.6)& (99.1; 116.0; 120.2)& (24.7; 29.5; 30.7) &(1.8; 2.2; 2.3)\\
xylophone &(45.8; 54.9; 57.2)&(136.7; 160.8; 166.8)&(37.8; 45.4; 47.3)&(3.0; 3.6; 3.8)\\
\\
\hline
\\
TOTAL & & & & \\
initial design &(30.8; 36.7; 38.2)& (100.4; 117.4; 121.7)&(25.0; 29.8; 31.0)&(1.9; 2.3; 2.4)\\
xylophone &(46.9; 56.2; 58.5)&(139.5; 164.2; 170.1)&(38.5; 46.2; 48.1)&(3.3; 4.0; 4.1)\\
\\
\hline

\end{tabular}\\
\end{center}
\end{table*}

\begin{table*}[ht]
\footnotesize 
\caption{Expected numbers of lensed GW events from inspiralling DCOs of different classes under different evolutionary scenarios. ``Low-end'' metallicity evolution assumed.
Predictions for the Einstein Telescope in the initial and ``xylophone'' configuration.}
\label{lensing-low}
\begin{center}  
\begin{tabular}{cccccc}

\hline
ET configuration  &standard&optimistic&delayed SN& high BH kicks  \\
$T_{surv}$ &{\scriptsize (1yr; 5yrs;continuous)}&{\scriptsize (1yr; 5yrs;continuous)}
&{\scriptsize (1yr; 5yrs;continuous)}&{\scriptsize (1yr; 5yrs;continuous)}\\
\hline\\

NS-NS & & & & \\
initial design &(0.02; 0.02; 0.02)&(0.2; 0.2; 0.2) &(0.03; 0.03; 0.03)&(0.02; 0.03; 0.03)\\
xylophone &(0.07; 0.07; 0.08)&(0.8; 0.9; 0.9)&(0.08; 0.09; 0.09)&(0.07; 0.08; 0.08)\\
\\
\hline\\

BH-NS & & & & \\
initial design &(0.7; 0.8; 0.8)&(1.3; 1.5; 1.5)&(0.4; 0.4; 0.4)&(0.07; 0.08; 0.08)\\
xylophone &(1.3; 1.4; 1.5)&(2.3; 2.7; 2.7)&(0.7; 0.8; 0.8) &(0.1; 0.2; 0.2)\\
\\
\hline\\

BH-BH & & & & \\
initial design &(39.2; 46.3; 48.1)&(110.7; 128.8; 133.4)&(32.1; 38.1; 39.6)&(2.4; 3.0; 3.0)\\
xylophone &(57.1; 68.0; 70.6)&(149.1; 174.4; 180.7)&(47.5; 56.7; 59.0)&(3.8; 4.6; 4.8)\\
\\
\hline\\

TOTAL & & & & \\
initial design &(39.9; 47.1; 48.9)&(112.3; 130.5; 135.1)&(32.5; 38.6; 40.1)&(2.5; 3.0; 3.2)\\
xylophone &(58.4; 69.4; 72.2)&(152.3; 178.0; 184.4)&(48.3; 57.6; 59.9)&(4.5; 4.9; 5.1)\\
\\
\hline\\

\end{tabular}\\
\end{center}
\end{table*}


\begin{table*}[ht]
\caption{Dependence of the expected number of lensed GW events per year on the velocity distribution function of lensing galaxies. For comparison, the values of parameters were taken from \citet{Mitchell2005}, \citet{Choi2007} and \citet{Bernardi2010}.
Predictions are for the Einstein Telescope in the initial and ``xylophone'' configuration. Long-term (``continuous'') operation assumed. }
\label{VDF}
\begin{center}  
\begin{tabular}{cccc}

\hline \
Velocity Distribution Function & Mitchell & Choi & Bernardi \\

\hline \\

 NS-NS systems &&& \\
initial design &&& \\
standard scenario & 0.05 & 0.07 & 0.09 \\
optimistic OCE & 0.2 & 0.2 & 0.3 \\
\\
``xylophone'' design &&& \\
standard scenario & 0.1 & 0.2 & 0.2 \\
optimistic OCE & 0.6 & 0.8 & 1. \\
\\
\hline \\

 BH-NS systems &&& \\
initial design &&& \\
standard scenario & 0.4 & 0.5 & 0.7 \\
optimistic CE scenario & 0.9 & 1.3 & 1.7 \\
\\
``xylophone'' design &&& \\
standard scenario & 0.8 & 1.1 & 1.4 \\
optimistic CE scenario & 1.8 & 2.5 & 3.2 \\
\\
\hline \\

BH-BH systems &&& \\
initial design &&& \\
standard scenario & 26.5 & 37.6 & 48.2 \\
optimistic CE scenario & 84.8 & 120.2 & 153.8 \\
\\
``xylophone'' design &&& \\
standard scenario & 40.3 & 57.2 & 73.2 \\
optimistic CE scenario & 117.6 & 166.8 & 213.6 \\
\\
\hline

\end{tabular}\\
\end{center}
\end{table*}

Expected yearly detection rates of DCO inspiralling systems are shown on Figure~\ref{rates} and reported in details in Table~\ref{rates-hi} (for the ``high-end'' metallicity evolution) and Table~\ref{rates-low} (for the ``low-end'' metallicity evolution). Figure~\ref{diff lensing} shows differential lensing probabilities $ \frac{1}{{\dot N}_{lensed}} \frac{d{{\dot N}_{lensed}}}{dz}.$ As compared with the probability density of DCO inspiral events shown on Figure~\ref{diff_rate}, now the distributions are shifted to higher redshifts.
On Figure~\ref{lensing rates} we display the yearly detection of lensing events for different classes of DCO systems. Horizontal dash-dotted line shows the threshold of one lensing event detected per year.
The rate of lensed inspiral events (shown here only for the ``low-end'' metallicity evolution) turns out to be noticeable for the BH-BH systems only.
In the next set of tables: Table~\ref{lensing-hi},\ref{lensing-low}, one can see the predicted yearly rates of gravitationally lensed DCO inspiral events to be seen by the Einstein Telescope. According to the discussion at the end of Section~\ref{sec:methodology} for each DCO evolutionary scenario and ET design we report three numbers corresponding respectively to 1 year, 5 years duration and continuous operation of the ET. Since the ET is planned to operate for a long time, the most relevant prediction would be in the last item (continuous operation). The preceeding ones should be understood as estimates on the number of lensed events one might expect during first 1 or 5 years of successful operation. We have checked that already temporal horizon of 10 years give predictions indistinguishable from "continuous" mode. At last in Table~\ref{VDF} we presented the rates of lensed inspiral events for initial and ``xylophone'' design calculated for three different models of velocity distribution function i.e. with Schechter-like function parameters obtained by \citet{Mitchell2005} ($n_{*} = 4.1\;10^{-3} (H_0/100)^3 \; Mpc^{-3}$, $\sigma_{*} = 88.8\;km/s$, ${\alpha} = 6.5$, $\beta = 1.93$), \citet{Choi2007} -- the ones used throughout this paper ($n_{*} = 8.\;10^{-3} (H_0/100)^3 \; Mpc^{-3}$, $\sigma_{*} = 161\;km/s$, ${\alpha} = 2.32$, $\beta = 2.67$) and \citet{Bernardi2010} ($n_{*} = 2.611\;10^{-3} (H_0/70)^3 \; Mpc^{-3}$, $\sigma_{*} = 159.6\;km/s$, ${\alpha} = 0.41$, $\beta = 2.59$). The purpose of Table~\ref{VDF} is to show how much lensing rate predictions depend on the assumptions concerning the population of lenses.

\section{Results and conclusions} \label{sec:conclusions}

The last row in Tables presented in Section~\ref{sec:rates} contains the total prediction. This is the one reflecting what could be the observed rates. Other rows show the expected contributions from distinct types of DCO.
 It can be seen from Table~\ref{rates-hi} and Table~\ref{rates-low} that dominating contribution to the total detection rate of DCO inspirals comes from BH-BH systems they contribute ca. $91 - 95 \%$ depending on the metallicity evolution scenario and the ET configuration in the standard scenario of DCO formation.
 Accordingly, the percentage of NS-NS events ranges from $1\%$ to $4\%$. This trend is present in all alternative DCO formation scenarios except the ``high-kick BH'' scenario in which BH-BH systems contribute only by $50-55\%$ assuming ``high-end'' metallicity evolution scenario and $72-75\%$ within the ``low-end'' scenario. This result can be understood since high-kick BH scenario tends to disrupt systems preventing the formation of DCOs containing the BH. So it is no surprise that NS-NS systems can contribute as much as $2-5\%$ within this scenario. The predictions concerning lensed events (continuous -- i.e. average yearly detection rate in c.a. 10 years of operation) comprise: $38-59$ detections per year in the standard scenario (higher end refers to the final ``xylophone'' design), which can be as high as $122-172\;[yr^{-1}]$ for the OCE and as low as $2-4\;[yr^{-1}]$ for high BH kicks scenario. This is for the ``high-end'' metallicity evolution. For the ``low-end'' scenario NS-NS contribute less than $1\%$ (actually less than $0.2\%$) so they would be unlikely to show up during the decades of operation. Consequently lensed BH-BH systems would be registered $ 98\%$ of the time. In absolute numbers, lensed NS-NS binaries are expected to be registered at rate ca. $2-7$ per century, only OCE scenario predicts ca. $2$ per decade. This means that if lensed inspiral event will be seen at the ET it will most likely come from BH-BH inspirals. In such case we can reasonably expect ca. $38 - 60$ lensed inspiral events per year, in the OCE it would be much higher ca. $150$ events per year. Only high BH kicks scenario predicts about a few events per year. In summary, our study based on the population synthesis evolutionary calculations of \citep{BelczynskiII} demonstrated that ET will likely provide us with a considerable statistics of lensed inspiral events.
 Such strongly lensed GWs from DCO inspirals would be dominated by BH-BH binaries, so it would be beneficial to gain more detailed knowledge about progenitors of such sources and their intrinsic mass distribution. In this context our results are also conservative since we assumed a point value for the chirp mass representative of the average value, but in fact BH-BH chirp mass distribution is known to be wide \cite{BelczynskiI,BelczynskiII}.

 Let us return to the question of how our predictions depend on the assumed form of the velocity distribution function for lensing galaxies. From Table~\ref{VDF} one can see that the differences are very small for NS-NS and BH-NS systems, but for BH-BH systems the differences are noticeable (up to a factor of two). Because BH-BH binaries are dominant sources, this means that with future progress on the galaxy velocity distribution function acquired with subsequent SDSS data releases or other large scale projects the inspiral DCO lensing rates should be reassessed. 

 The discovery of strong lensing inspiral signal would most likely be {\it ex post} by searching for signals with identical frequency drift and wave strain   pattern differing from each other only by amplitudes. This is quite different from the optical transient sources like supernovae or gamma-ray bursts which can be regarded as standard (or rather standardizable) candles. For them registering an unusually bright event would be the trigger and most likely value of time delay would suggest when and where to search for the fainter image --- in optical studies we should know where to point the telescope. On the other hand, the gravitational wave detectors essentially see the whole sky (up to the antenna pattern setting the sky coverage). Hence in our case the strategy would be different and statistical description of time delays and magnifications would be of little value. Let us remark that NS-NS systems could be considered as standardizable sirens (due to narrow chirp mass distribution) and for them one can think of using magnification and time delay distributions.
However, as we see they contribute negligibly to the total rate of lensed events. The dominant sources --- BH-BH DCO systems ---  have very wide chirp mass distribution and consequently the strategy to detect lensed events should be {\it ex post}.

 The benefits from detecting lensed inspiral events are manifold. It is well known, that from a single inspiral event we would be able to derive redshifted chirp mass $(1+z_s){\cal M}_c$ and luminosity distance $d_L(z_s)$. These quantities are determined by combining temporal pattern of the amplitude $h(t)$ and the frequency drift ${\dot f(t)}$. In the case of multiple images we will have multiple signals (each from the same $d_L(z_s)$ and having the same $(1+z_s){\cal M}_c$) whose amplitude is scaled according to the image magnification, which depends on single parameter -- the lens-source misalignment $y$.
 As pointed out by \citet{Sereno2}, who first discussed similar problems in the context of LISA, we would be able to determine $y$ from image amplitude ratios.
 These amplitude ratios are GW analogs of flux ratios known in strong lensing. Flux ratios, however are sensitive to differential extinction of light passing through different parts of lensing galaxy. This would not be a problem for GWs. Moreover, if the lensing galaxy could be identified in the optical as a strong lensing system, then comparative analysis of GW amplitude ratios and optical flux ratios would create a unique possibility to study galactic structure. Such optical lens identification would be extremely valuable: from GWs we would have accurate time delays, the $y$ parameter (one of severe confounders in modelling strong lenses from time delays) and from the optical we would have image astrometry and lens spectroscopy. The benefits of having all this together are obvious. So one can imagine that in the future lensed GW inspiral signals would even trigger dedicated deep surveys seeking for strong lensing systems.

\section*{Acknowledgments}
The authors are grateful to the referee for very useful comments which allowed to improve the paper. 
M.B. is supported by the Recruitment Program of High-end Foreign Experts and he gratefully acknowledges hospitality of Beijing Normal University where this research was finished. A.P. and M.B. are partly supported by the Poland-China Scientific \& Technological Cooperation Committee Project No. 35-4.
Z.-H.Z. is supported by the Ministry of Science and Technology National Basic Science Program (Project 973) under Grants Nos. 2012CB821804 and 2014CB845806, the National Natural Science Foundation of China under Grants Nos. 11373014 and 11073005, and the Fundamental Research Funds for the Central Universities and Scientific Research Foundation of Beijing Normal University.


\end{document}